\documentclass{article}
\usepackage{spconf,amsmath, graphicx}
\usepackage{diagbox}
\usepackage{cite}
\usepackage{booktabs}
\usepackage{array}
\usepackage{multirow}
\usepackage{caption}
\usepackage{subcaption}
\usepackage[strings]{underscore} 

\title{RoboVox Far Field Speaker Recognition: A Novel Data Augmentation Approach with Pretrained Models}

\name{Muhammad Sudipto Siam Dip, Md Anik Hasan, Sapnil Sarker Bipro, Md Abdur Raiyan, Mohammod Abdul Motin}
\address{Department of Electrical \& Electronic Engineering,  Rajshahi University of Engineering \& Technology}

\begin{document}
\maketitle
\begin{abstract}
In this study, we address the challenge of speaker recognition using a novel data augmentation technique of adding noise to enrollment files. This technique efficiently aligns the sources of test and enrollment files, enhancing comparability. Various pre-trained models were employed, with the resnet model achieving the highest DCF of 0.84 and an EER of 13.44. The augmentation technique notably improved these results to 0.75 DCF and 12.79 EER for the resnet model. Comparative analysis revealed the superiority of resnet over models such as ECPA, Mel-spectrogram, Payonnet, and Titanet large. Results, along with different augmentation schemes, contribute to the success of RoboVox far field speaker recognition in this paper.
\end{abstract}

\begin{keywords}
speech augmentation, far-field speaker recognition, pre-trained model
\end{keywords}

\section{Introduction}
\label{sec:intro}
Speaker recognition systems are extensively utilized in applications related to home customization, authentication, and security. It is a biometric technology that helps to identify the system whether a pair of utterances correlate to the same speaker or not. The speaker verification generally contains a speaker pulling out embedding and a scoring process. During the embedding extraction process, audio of varying lengths is transformed into one fixed-dimensional vector representation known as a speaker embedding. This embedding is intended to include information about the speaker. For the scoring method, cosine similarity or Euclidean distance can be used. In recent years, with the development of computing power, deep learning techniques have been popular for the speaker verification process. However, the effectiveness significantly decreases when the speech is obtained in natural, uncontrolled environments such as far-field noisy environments with variable distance and reverberation. There are some benchmarks for these problems are VoiCes and FFSVC.\cite{Nagrani19}\cite{Chung18b} However, they don’t include the internal noise of the device and the angle between the device and the speakers.

Usually, voice samples collected from distances are rather small and not enough to build high-quality speaker verification models without any prior training. Consequently, near-field datasets are generally utilized for training to enhance the classification performance of speaker verification systems. There are different kinds of transfer learning methods used to address this domain mismatch. Data augmentation is the most used technique to address this domain mismatch and train a  robust neural network. Simulated reverberation\cite{Liu2023}, additive noise, and Specaugment\cite{SpecAugment} are effective methods for augmenting data in speaker verification. These techniques can expand the variety of acoustic environments that may be encountered during real-life scenarios.

The challenge of SPCUP-2024 is Robovox: far-field speaker recognition by a mobile robot. We started the quest to solve this challenge by extracting embeddings of raw signals through neural networks. After going through several experiments we have that the system works well with pre-trained models and even better if we augment the near clean train data with some artificial noise. Finally, we have developed a novel augmentation method that helped us to attain optimal performance in both the EER and DCF assessment criteria. Our noise and reverberation augmentation techniques for real-life scenarios surpass our different experimental approaches. The rest of the papers are organized as follows. Section 2 represents the methodology of our work with different augmentation techniques. The results of our experiment are shown in section 3, while 4 discusses the results based on our experiments. The conclusion is added in section 5.

\section{Methodology}
\label{sec:majhead}

\begin{figure*}[!t]
    \centering
    \includegraphics[width=.8\textwidth]{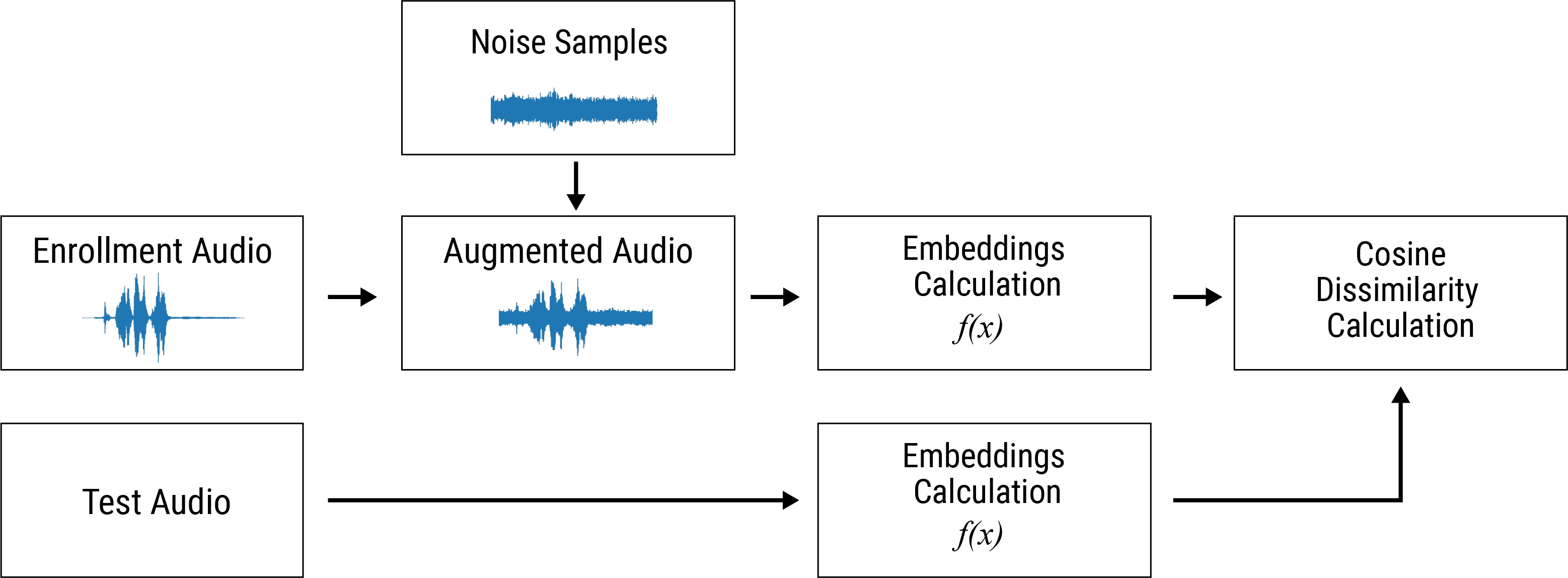}
    \caption{An overall representation of our implemented framework. It starts by taking audio files from channel 5 and mixing them with noise extracted from audio files related to channel 4. The augmented enrollment signal is used to calculate the vector embedding set with the help of a large deep-learning model. Simultaneously the audio files from the test set are also fetched to calculate the embedding. Both of these are compared with the cosine dissimilarity evaluation metric}
    \label{fig:overall-work}
\end{figure*}

\subsection{Dataset Description}
\label{ssec:subhead}
This competition leverages the Robovox dataset. This dataset adds a novel benchmark in the research of far-field single-channel and multi-channel speaker verification. A robot is equipped with three microphones positioned at the angle of the robot (channels 1 to 3). The fourth microphone (channel 4) is placed inside the robot. Another microphone (channel 5) used as a ground truth microphone is placed close to the speaker. The dataset comprises 2,219 conversations spoken by 78 individuals. Each conversation is composed of an average of 5 dialogues, resulting in a total of around 11,000 dialogues. The average recording of the dialogue is 3.5 seconds. The dataset was recorded at different distances of 1m, 2m, and 3m from the speakers. To emulate real-life scenarios the session is recorded in different room environments in the hall, open space, and small and medium rooms while placing the robot at the wall, center, and corner. The dataset contains two parts for single-channel and multichannel tracks. For this competition, a single channel is utilized. For the enrollment files channel 5 and the test files, channel 4 is chosen, containing 225 and 10,332 files consecutively. 

\subsection{Preprocessing}
\label{ssec:subhead}
In the conventional machine learning approach, the data the model is trained on and evaluated needs to come from the same source. If the audio enrollment that is used for the learning algorithm is from one source and then the data is evaluated from data of another source raises difficulty. In this benchmark, the enrollment files are recorded with channel 5, which is the best channel. On the other hand, the test dataset is recorded with channel 4 which is the most challenging channel. The enrollment audio files are less noisy and ambient than test audio files as they were recorded closer to the speaker. The test audio is not only noisy but contains multiple variabilities with that noise such as reverberations, angles etc. Thus, the signal-to-noise (SNR) ratio from those two sources contains massive dissimilarity. Another problem was, that there was no voice activity in some audio files provided such as spk{\_}6-6{\_}11{\_}0{\_}0{\_}d4{\_}ch5 and all the files for spk{\_}21 and some other files as well in enrollment and test dataset. Assigning subject labels to these data and then feeding it into the learning algorithm alongside other data risks misleading the algorithm and preventing it from learning correctly. Thus, instead of focusing on improving the learning algorithms itself, we focused more on how the data set could be improved which can be later used for feature extraction. We applied two different schemes that can reduce the mismatch between these two sources and improve proximity. In our first scheme, we focused on reducing noise from the test data upon simulating the noise reduction on enrollment data. In our second scheme, we focused on augmenting the enrollment files with similar and equivalent noise to the test dataset with our developed approach.

\subsubsection{Noise Reduction}
\label{sssec:subsubhead}
In this scheme, we tried to reduce the noise from the test dataset before it was fetched into a deep learning model for embedding calculation. We used the noisereduce library \cite{noiseReducePaper} which is a common library to reduce noise for stationary and non-stationary signals. We set the parameter of the proportion to reduce the noise by 100{\%} and threshold for non-stationary noise reduction 1 considering the test files containing different variabilities.

\subsubsection{Data augmentation with noise samples}
\label{sssec:subsubhead}
In our second scheme of preprocessing, we utilized data augmentation by adding noises to our enrollment audio files. In general, two approaches can be used. First, noise such as Gaussian noise, pink noise, etc. can be simulated by different available libraries namely numpy, pudub etc. Secondly, background noises can collected from different resources and datasets such as AudioSet \cite{AudioSet}.

\begin{figure}[!t]
    \centering
    \begin{subfigure}[b]{0.5\textwidth}
        \includegraphics[width=\textwidth]{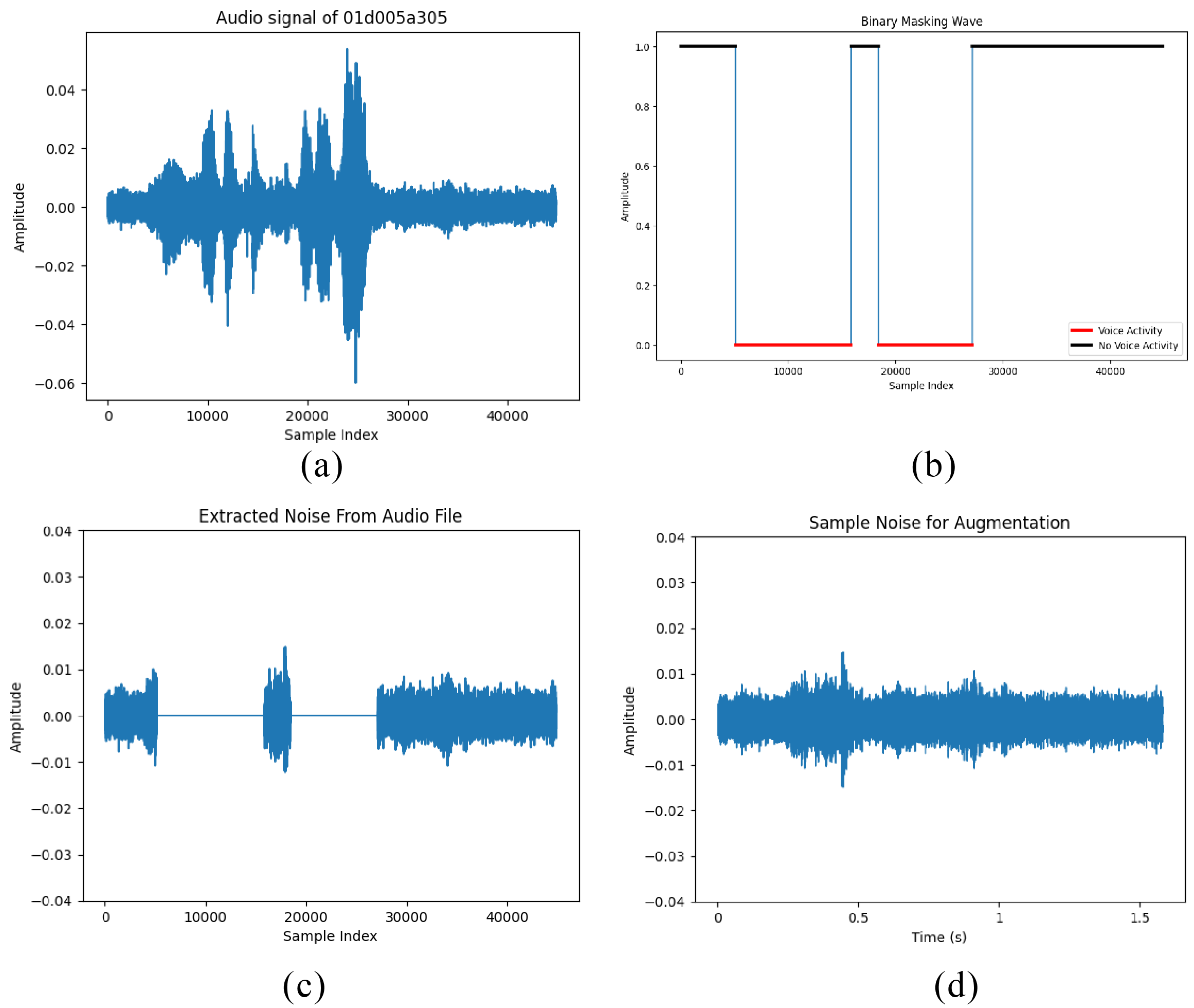}
    \end{subfigure}
    \hfill
    
    \caption{The step-by-step response for the implemented noise extraction algorithm. Where a) An audio file recorded with microphone 4, b) Binary mask obtained by determining voice activity intervals c) Result of multiplication binary mask with the audio signal d) Keeping non-zero segments i.e. noise.}
    \label{fig:step-by-step-response}
\end{figure}

We hypothesize that reducing the mismatch between two sources may lead to reducing the cosine dissimilarity for the same speaker and increasing dissimilarity between different speakers. Instead of generating noise from different resources we simulated the noises from audio files which were recorded with channel 4 and used them to augment the enrollment dataset which was recorded with channel 5. We start by setting a threshold by manually inspecting the audio files and then detect the voice activity intervals where the amplitude in decibels is above the threshold. The intervals are used to create a binary masking which is multiplied by the corresponding sample audio signal to extract noise samples shown in figure \ref{fig:step-by-step-response} which were used for augmenting audio files from channel 5. The step-by-step process is shown in the algorithm table.

\begin{table}[!ht]
\begin{tabular}{clll}
\hline
\multicolumn{4}{l}{\textbf{\begin{tabular}[c]{@{}l@{}}Algorithm: Noise extraction from audio files for \\                     augmentation\end{tabular}}} \\ \hline
1 & Start & \multicolumn{1}{c}{} & \multicolumn{1}{c}{} \\
2 & \multicolumn{3}{l}{Input: Audio signal, Threshold in dB} \\
3 & \multicolumn{3}{l}{For each sample in an audio signal:} \\
4 &  & \multicolumn{2}{l}{if sample in dB \textless threshold:} \\
5 &  &  & Start of non-silent period \\
6 &  & \multicolumn{2}{l}{Else} \\
7 &  &  & End of non-silent period \\
8 & \multicolumn{3}{l}{\begin{tabular}[c]{@{}l@{}}Make a list of ranges of non-silent periods\\ in an audio signal.\end{tabular}} \\
9 & \multicolumn{3}{l}{\begin{tabular}[c]{@{}l@{}}Initialize an empty list to store \\ expanded non-silent indices\end{tabular}} \\
10 & \multicolumn{3}{l}{For each interval in non-silent intervals:} \\
11 &  & \multicolumn{2}{l}{Expand the interval to individual   indices.} \\
12 &  & \multicolumn{2}{l}{\begin{tabular}[c]{@{}l@{}}Add the expanded indices to the list \\ of non-silent   indices\end{tabular}} \\
13 & \multicolumn{3}{l}{For each index in the list of non-silent indices:} \\
14 &  & \multicolumn{2}{l}{Set the corresponding index in the mask to 0} \\
15 & \multicolumn{3}{l}{\begin{tabular}[c]{@{}l@{}}Multiply the mask with the original \\ signal (element-wise multiplication)\end{tabular}} \\
16 & \multicolumn{3}{l}{\begin{tabular}[c]{@{}l@{}}The result is the noise-only signal, \\ where non-silent parts are nullified\end{tabular}} \\
17 & \multicolumn{3}{l}{Return the noise-only signal} \\
18 & \multicolumn{3}{l}{End} \\ \hline
\end{tabular}
\end{table}

\begin{table}[]
\caption{Comparison of results before augmentation}
\centering
\begin{tabular}{ccc}
\hline
\textbf{Models}       & \textbf{EER} & \textbf{DCF} \\ \hline
Pyannote              & 19.84        & 0.99         \\
ECAPA-TDNN            & 15.59        & 0.89         \\
Titanet Large         & 15.75        & 0.88         \\
Mel-Spec + ECAPA-TDNN & 15.05        & 0.88         \\
ResNet-TDNN           & 13.44        & 0.84         \\ \hline
\end{tabular}
\label{tab:before-augmentation}
\end{table}

\begin{table*}[!t]
\begin{center}
\caption{Comparison of results after noise reduction and augmentation. Three augmentation schemas have been used with different levels of signal-to-noise ratio (SNR).}
\begin{tabular}{|c|cc|cccccc|}
\hline
\multirow{3}{*}{Feature extractor models} & \multicolumn{2}{c|}{\multirow{2}{*}{Noise Reduction}} & \multicolumn{6}{c|}{Augmentation} \\ \cline{4-9} 
 & \multicolumn{2}{c|}{} & \multicolumn{2}{c|}{\begin{tabular}[c]{@{}c@{}}Augmentation 1\\ SNR(-10dB to -4dB)\end{tabular}} & \multicolumn{2}{c|}{\begin{tabular}[c]{@{}c@{}}Augmentation 2\\ SNR(-7dB to -4dB\end{tabular}} & \multicolumn{2}{c|}{\begin{tabular}[c]{@{}c@{}}Augmentation 3\\ SNR(-7dB to +3dB)\end{tabular}} \\ \cline{2-9} 
 & \multicolumn{1}{c|}{EER} & DCF & \multicolumn{1}{c|}{EER} & \multicolumn{1}{c|}{DCF} & \multicolumn{1}{c|}{EER} & \multicolumn{1}{c|}{DCF} & \multicolumn{1}{c|}{EER} & DCF \\ \hline
ECAPA-TDNN & \multicolumn{1}{c|}{19.53} & 0.96 & \multicolumn{1}{c|}{23.25} & \multicolumn{1}{c|}{0.9} & \multicolumn{1}{c|}{24.02} & \multicolumn{1}{c|}{0.91} & \multicolumn{1}{c|}{25.57} & 0.97 \\ \hline
Mel-Spectrogram + ECAPA-TDNN & \multicolumn{1}{c|}{18.56} & 0.94 & \multicolumn{1}{c|}{14.47} & \multicolumn{1}{c|}{0.85} & \multicolumn{1}{c|}{14.66} & \multicolumn{1}{c|}{0.83} & \multicolumn{1}{c|}{14.47} & 0.85 \\ \hline
Resnet-TDNN & \multicolumn{1}{c|}{18.15} & 0.93 & \multicolumn{1}{c|}{\textbf{12.72}} & \multicolumn{1}{c|}{\textbf{0.75}} & \multicolumn{1}{c|}{12.99} & \multicolumn{1}{c|}{77} & \multicolumn{1}{c|}{13.63} & 0.77 \\ \hline
\end{tabular}
\label{tab:after-augmentation}
\end{center}
\end{table*}

\subsection{Extracting the embedding}
\label{sec:print}
After the preprocessing, we extracted the embedding using 4 pre-trained models of different architectures. The ECAPA-TDNN \cite{ecapa} is a Time Delay Neural Network (TDNN) based model built upon x-vector architecture and using multi-scale Res2Net features. This speaker-embedding extractor model emphasizes Channel Attention, Propagation, and Aggregation. The pre-trained model used in our implementation is trained with Voxceleb1\cite{Nagrani19} and Voxceleb2 \cite{Chung18b} training data. We have also tested the model with mel-spectrogram as input instead of using direct raw audio. The Resnet-Tdnn \cite{resnet} model is based on 34 layered residual networks. While the pyannote audio toolbox\cite{pyannote} has used a cannocial x-vector-based TDNN-based architecture. Also, SincNet\cite{sincNet} features have been used in this pre-trained model. Finally, the titanet\cite{titanet} has used Squeeze-and-Excitation layers followed by channel attention-based statistics pooling layer. All of the models were used from the speechbrain package \cite{speechbrain}.

\section{Result}
\label{sec:result}

For speaker embedding calculation five different pretrained models (Paynnote, ECPA, Titanet, Melspectrogram, and ResNet) were employed. To address the challenge that, multiple files associated with a single speaker, the average embedding was computed to derive the final representation of each speaker. The results of each model are summarized in Table \ref{tab:before-augmentation}. Paynnote exhibited a DCF of 0.99 and an EER of 19.85, while ECAPA outperformed Paynnote with a DCF of 0.89 and an EER of 15.59. Titanet and Melspectrogram demonstrated further improvements, achieving DCF values of 0.88, with corresponding EER results of 15.75 and 15.8, respectively. The ResNet model attained the best results, with a DCF of 0.84 and an EER of 13.44.
The evaluation extended to the impact of noise reduction of test files and different enrollment file augmentation processes on the performance of ECPA, ResNet and Melspectrogram models, detailed in Table 2. The noise reduction applied to the test files appeared to have limited effectiveness, as the models yielded higher DCF and EER values compared to the original test file's dissimilarity. Specifically, ECPA, Melspectrogram, and ResNet attained DCF values of 0.96, 0.94, and 0.93, respectively. Employing three distinct frequency ranges for data augmentation (-3 dB to -17 dB, -7 dB to -4 dB, and -10 dB to -4 dB), we observed notable enhancements. In the first augmentation process (Aug3), both ResNet and Melspectrogram outperformed the models using only original enrollment files. Specifically, Melspectrogram improved from a DCF of 0.88 to 0.85, and ResNet improved from 0.84 to 0.77. The second frequency range (-7 dB to -4 dB) proved effective for both models, resulting in improved DCF values of 0.83 and 0.77 for Melspectrogram and ResNet, respectively. The third augmentation process, based on the frequency range of -10 dB to -4 dB, emerged as the optimal solution for both Melspectrogram and ResNet, achieving DCF values of 0.82 and 0.75, respectively. Conversely, the ECAPA model exhibited DCF values of 0.99, 0.91, and 0.90 for these three augmentation processes, indicating a sensitivity to variations in the augmentation parameters.

\section{Discussion}
\label{sec:page}

These findings contribute substantial insights to the field of speaker verification, specifically within the context of the Robovox far field speaker recognition dataset. The nuanced performance variations observed across pretrained models underscore the critical importance of carefully selecting model architectures tailored to the characteristics of the dataset and specific application requirements. Notably, the effectiveness of ResNet, especially when coupled with different augmentation strategies, suggests its robustness in capturing the unique speaker characteristics prevalent in the Robovox far field dataset. Conversely, the observed sensitivity of the ECAPA model to augmentation parameters emphasizes the need for model-specific optimization, particularly when applied to this distinct dataset.
Furthermore, the study shows the potential of data augmentation technique by incorporating noise in elevating the overall performance of speaker verification systems, specifically in the context of the Robovox far field dataset. The discernible improvements in DCF values and EER across different augmentation processes underscore the paramount importance of tailoring augmentation strategies to the intricacies of the Robovox far field dataset and the pretrained models employed.
The study not only advances our understanding of pretrained model performances but also highlights the significance of dataset-specific considerations and augmentation strategies in optimizing speaker verification outcomes within the unique characteristics of the Robovox far field dataset.

\section{Conclusion}
\label{sec:illust}
In this paper, a novel data augmentation technique, noise addition to enrollment files was employed and, the resnet pretrained model's notable effectiveness, achieving a DCF of 0.75 and an EER of 12.79 was observed. The application of our data augmentation technique significantly improved the model's performance, reducing the DCF rating from 0.84 to 0.75. This indicates the efficacy of the proposed approach, tailored to the characteristics of the Robovox far field speaker dataset, positioning the data augmentation technique as a valuable tool for addressing speaker verification challenges. The study establishes a concise yet impactful strategy for enhanced speaker recognition outcomes, contributing to advancements in far field speaker verification system.

\bibliographystyle{ieeetr}

\end{document}